\documentclass[runningheads]{llncs}
\usepackage{makecell}
\usepackage{graphicx}
\usepackage{amssymb}
\usepackage{framed,multirow,color}
\usepackage{marvosym}
\usepackage{threeparttable}
\begin{document}
\title{Frequency-mixed Single-source Domain Generalization for Medical Image Segmentation}
\titlerunning{FreeSDG for Medical Image Segmentation}
\author{Heng Li\inst{1,3} \and 
Haojin Li\inst{1,2} \and 
Wei Zhao\inst{3,4,}\textsuperscript{\Letter} \and 
Huazhu Fu\inst{5} \and 
Xiuyun Su\inst{3,4} \and 
Yan Hu\inst{2} \and 
Jiang Liu\inst{1,2,3,6,}\textsuperscript{\Letter}}
%
\authorrunning{Heng Li, Haojin Li, Wei Zhao et al.}
\institute{Research Institute of Trustworthy Autonomous Systems, Southern University of Science and Technology, Shenzhen, China\and
Department of Computer Science and Engineering, Southern University of Science and Technology, Shenzhen, China\and
Medical Intelligence and Innovation Academy, Southern University of Science and Technology, Shenzhen, China\and
Southern University of Science and Technology Hospital, Shenzhen, China\and
Institute of High Performance Computing, Agency for Science, Technology and Research, Singapore\and
Guangdong Provincial Key Laboratory of Brain-inspired Intelligent Computation, Southern University of Science and Technology, Shenzhen, China\\
\email{\textsuperscript{\Letter} Corresponding authors: zhaow3, liuj@sustech.edu.cn}
}

\maketitle              
\begin{abstract}
The annotation scarcity of medical image segmentation poses challenges in collecting sufficient training data for deep learning models. 
Specifically, models trained on limited data may not generalize well to other unseen data domains, resulting in a domain shift issue. 
Consequently, domain generalization (DG) is developed to boost the performance of segmentation models on unseen domains. 
However, the DG setup requires multiple source domains, which impedes the efficient deployment of segmentation algorithms in clinical scenarios.
To address this challenge and improve the segmentation model's generalizability, we propose a novel approach called the Frequency-mixed Single-source Domain Generalization method (FreeSDG). 
By analyzing the frequency's effect on domain discrepancy, FreeSDG leverages a mixed frequency spectrum to augment the single-source domain. Additionally, self-supervision is constructed in the domain augmentation to learn robust context-aware representations for the segmentation task.
Experimental results on five datasets of three modalities demonstrate the effectiveness of the proposed algorithm. FreeSDG outperforms state-of-the-art methods and significantly improves the segmentation model's generalizability. 
Therefore, FreeSDG provides a promising solution for enhancing the generalization of medical image segmentation models, especially when annotated data is scarce.
The code is available at https://github.com/liamheng/Non- IID\_Medical\_Image\_Segmentation.
 
\end{abstract}

\section{Introduction}
Due to the superiority in image representation, tremendous success has been achieved in medical image segmentation through recent advancements of deep learning~\cite{qiu2022fgam}.
Nevertheless, sufficient labeled training data is necessary for deep learning to learn state-of-the-art segmentation networks, resulting in the burden of costly and labor-intensive pixel-accurate annotations~\cite{jiang2022multi}.
Consequently, annotation scarcity has become a pervasive bottleneck for clinically deploying deep networks, and existing similar datasets have been resorted to alleviate the annotation burden.
However, networks trained on a single-source dataset may suffer performance dropping when applied to clinical datasets, since neural networks are sensitive to domain shifts.

Consequently, domain adaptation (DA) and DG~\cite{zhou2022domain} have been leveraged to mitigate the impact of domain shifts between source and target domains/datasets.
Unfortunately, DA relies on a strong assumption that source and target data are simultaneously accessible~\cite{li2022annotation}, which does not always hold in practice.
Thereby, DG has been introduced to overcome the absence of target data, which learns a robust model from distinct source domains to generalize to any target domain.
To efficiently transfer domain knowledge across various source domains, FACT~\cite{QinweiXu2021AFF} has been designed to adapt the domains by swapping the low-frequency spectrum of one with the other.
Considering privacy protection in medical scenarios, federated learning and continuous frequency space interpolation were combined to achieve DG on medical image segmentation~\cite{liu2021feddg}.
More recently, single-source domain generalization (SDG)~\cite{peng2022out} has been proposed to implement DG without accessing multi-source domains.
Based on global intensity non-linear augmentation (GIN) and interventional pseudocorrelation augmentation (IPA), a causality-inspired SDG was designed in~\cite{ouyang2022causality}.
Although DG has boosted the clinical practice of deep neural networks, troublesome challenges still remain in clinical deployment.
1) Data from multi-source domains are commonly required to implement DG, which is costly and even impractical to collect in clinics.
2) Medical data sharing is highly concerned, accessing multi-source domains exacerbates the risk of data breaching. 
3) Additional generative networks may constrain algorithms' efficiency and versatility, negatively impacting clinical deployment.  

To circumvent the above challenges, a frequency-mixed single-source domain generalization strategy, called FreeSDG, is proposed in this paper to learn generalizable segmentation models from a single-source domain.
Specifically, the impact of frequency on domain discrepancy is first explored to test our hypotheses on domain augmentation.
Then based on the hypotheses, diverse frequency views are extracted from medical images and mixed to augment the single-source domain.
Simultaneously, a self-supervised task is posed from frequency views to learn robust context-aware representations.
Such that the representations are injected into the vanilla segmentation task to train segmentation networks for out-of-domain inference.
Our main contributions are summarised as follows:
\begin{itemize}
    \item We design an efficient SDG algorithm named FreeSDG for medical image segmentation by exploring the impact of frequency on domain discrepancy and mixing frequency views for domain augmentation.
    \item Through identifying the frequency factor for domain discrepancy, a frequency-mixed domain augmentation (FMAug) is proposed to extend the margin of the single-source domain.
    \item A self-supervised task is tailored with FMAug to learn robust context-aware representations, which are injected into the segmentation task. 
    
    \item Experiments on various medical image modalities demonstrate the effectiveness of the proposed approach, by which data dependency is alleviated and superior performance is presented when compared with state-of-the-art DG algorithms in medical image segmentation.
\end{itemize}

\begin{figure}[!t]
    \begin{centering}
        \includegraphics[width=1\linewidth]{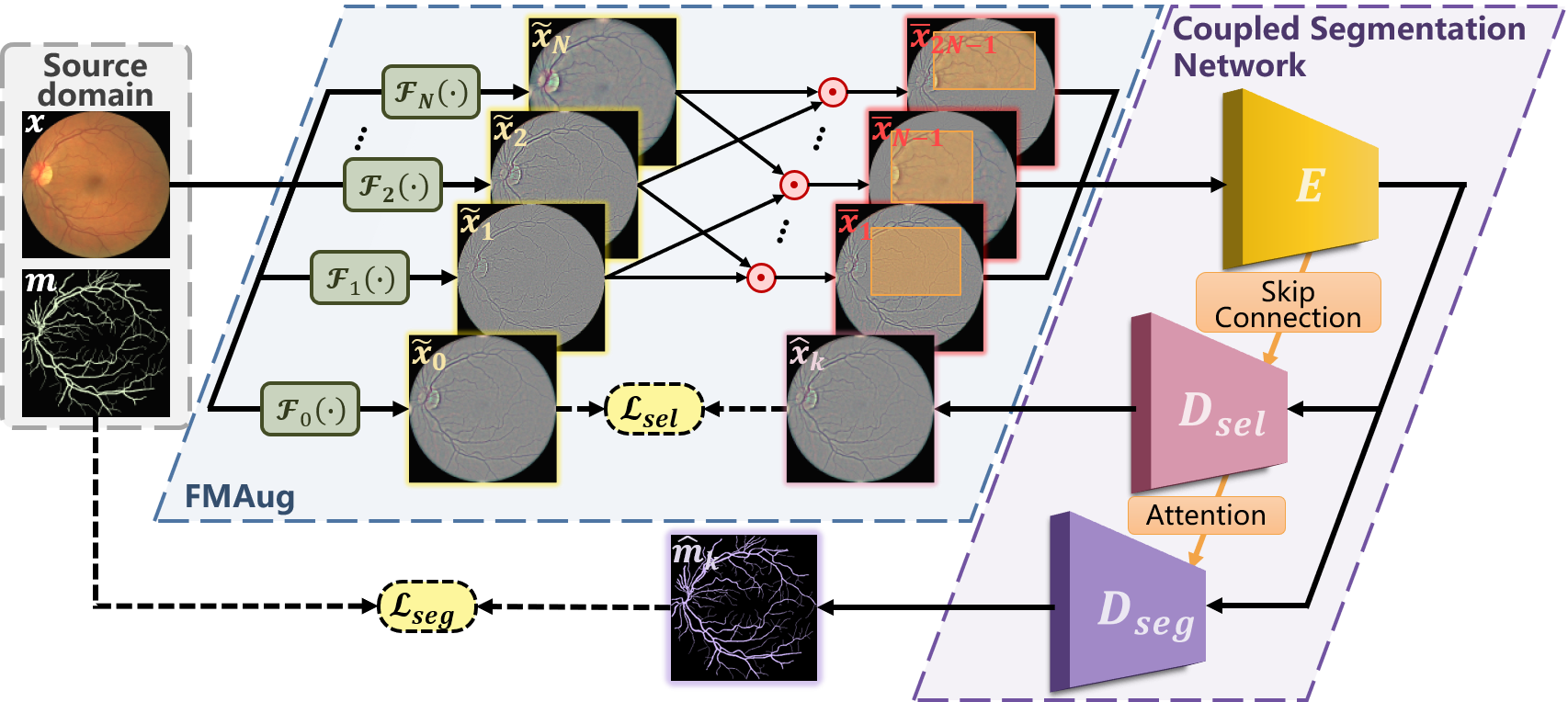}
        \par
    \end{centering}
\caption{Overview of FreeSDG, which learns a generalizable segmentation network from a single-source domain.
FMAug extends the domain margin by mixing patches (orange boxes) from diverse frequency views, and poses a self-supervised task to learn context-aware representations.
The representations are injected into segmentation using attention mechanisms in the coupled network to achieve a generalizable model. 
}
\vskip -5pt
\label{fig:workflow}
\end{figure}

\section{Methodology}
Aiming to robustly counter clinical data from unknown domains, an SDG algorithm for medical image segmentation is proposed, as shown in Fig.~\ref{fig:workflow}.
A generalizable segmentation network is attempted to be produced from a single-source domain $(x,m)\sim \mathbb{D}(x,m)$, where $m\in \mathbb{R}^{H\times W}$ is the segmentation mask for the image $x \in \mathbb{R}^{H\times W\times3}$.
By mixing frequency spectrums, FMAug is executed to augment the single-source domain, and self-supervision is simultaneously acquired to learn context-aware representations.
Thus a medical image segmentation network capable of out-of-domain generalization is implemented from a single-source domain.

\subsection{Frequency-controlled Domain Discrepancy} 
Generalizable algorithms have been developed using out-of-domain knowledge to circumvent the clinical performance dropping caused by domain shifts.
Nevertheless, extra data dependency is often inevitable in developing the generalizable algorithms, limiting their clinical deployment.
To alleviate the data dependency, a single source generalization strategy is designed inspired by the Fourier domain adaption~\cite{yang2020fda} and generalization~\cite{QinweiXu2021AFF}.

According to~\cite{QinweiXu2021AFF,yang2020fda}, the domain shifts between the source and target could be reduced by swapping/integrating the low-frequency spectrum (LFS) of one with the other.
Thus we post two hypotheses: 

1) uniformly removing the LFS reduces inter- and inner-domain shifts;

2) discriminatively removing the LFS from a single domain increases inner-domain discrepancy.

\begin{figure}[!t]
    \begin{centering}
        \includegraphics[width=\linewidth]{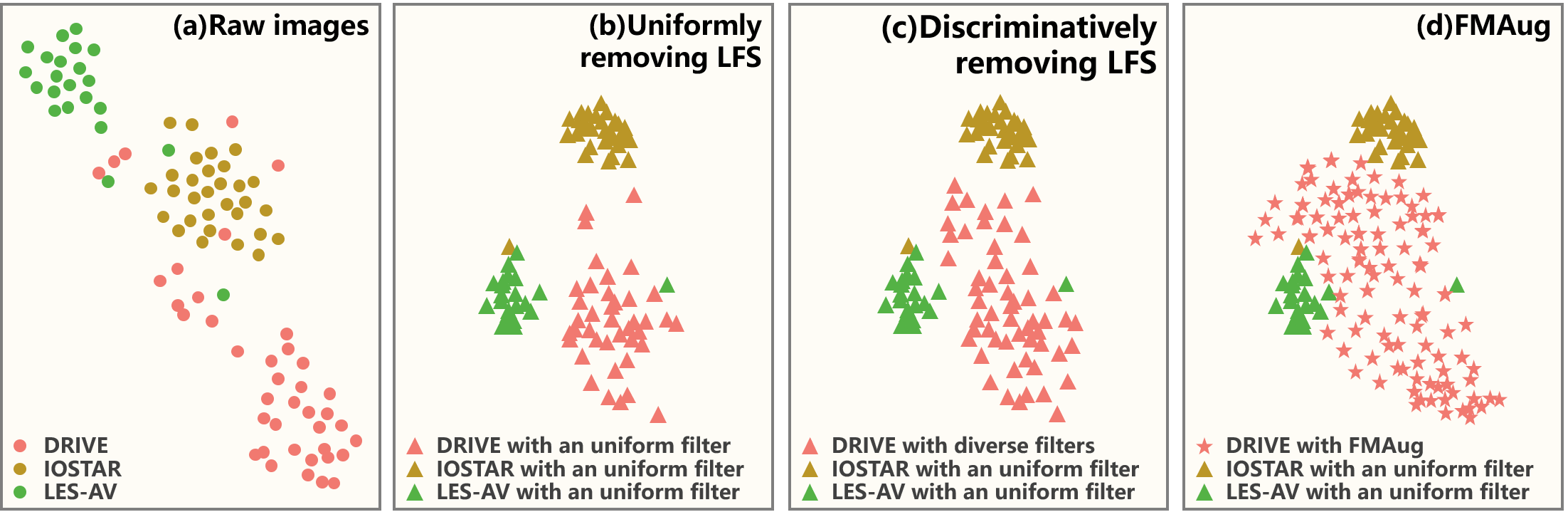}
        \par
    \end{centering}
\caption{Data distribution visualized by t-SNE. Uniformly removing LFS reduces shifts between DRIVE, IOSTAR, and LES-AV. Discriminatively removing the LFS increases the discrepancy in DRIVE.
FMAug extends the margin of DRIVE.} 
\vskip -5pt
\label{fig:t-SNE}
\end{figure}

Various frequency views are thus extracted from medical images with changing parameters to verify the above hypotheses.
Denote the frequency filter with parameters $\theta_n$ as $\mathcal{F}_n\left ( \cdot  \right )$, where $n \in \mathbb{R}^{N+1}$ refers to the index of parameters.
Following~\cite{li2022structure,li2022annotation}, a frequency view acquired with $\theta_n$ from an image $x$ is given by $\tilde{x}_n=  \mathcal{F}_n\left ( x  \right ) = x - x\ast g(r_n,\sigma_n)$, where $g(r_n,\sigma_n)$ denotes a Gaussian filter with radius $r_n \in [5,50]$ and spatial constant $\sigma_n \in [2,22]$.
Then the frequency views are converted to vectors by a pre-trained ResNet-18 and t-SNE is employed to demonstrate the domain discrepancy controlled by the low-frequency spectrum.

As shown in Fig.~\ref{fig:t-SNE}, compared to the raw images, the distribution of various datasets is more clustered after the uniform LFS removement, which indicates domain shift reduction.
While the domain discrepancy in DRIVE is increased by discriminatively removing the LFS.
Accordingly, these hypotheses can be leveraged to implement SDG.

\subsection{Frequency-mixed domain augmentation} 
Motivated by the hypotheses, domain augmentation is implemented by $\mathcal{F}_n\left ( \cdot  \right )$ with perturbed parameters. 
Moreover, the local-frequency-mix is executed to further extend the domain margin, as shown in Fig.~\ref{fig:t-SNE} (d).
As exhibited in the blue block of Fig.~\ref{fig:workflow}, random patches are cut from a frequency view and mixed with diverse ones to conduct FMAug, which is given by
\begin{equation}
\bar{x}_k= \mathcal{M}\left ( \tilde{x}_i,\tilde{x}_j \right ) = M\odot \tilde{x}_i+(1-M)\odot \tilde{x}_j, 
\label{eq:cutmix}
\end{equation}
where $M \in {0,1}^{W \times H}$ is a binary mask controlling where to drop out and fill in from two images,  and $\odot$ is element-wise multiplication. $k=(i-1) \times N + (j - 1)$ denotes the index of the augmentation outcomes, where $i,j \in \mathbb{R}^N, i\ne j$.

Notably, self-supervision is simultaneously acquired from FMAug, where only patches from $N$ frequency views $\tilde{x}_n, n \in \mathbb{R}^{N}$ are mixed, and the rest one $\tilde{x}_0$ is cast as a specific view to be reconstructed from the mixed ones, where $(r_n,\sigma_n)=(27,9)$.
Under the self-supervision, an objective function for learning context-aware representations from view reconstruction is defined as
\begin{equation}
\mathcal{L}_{sel}=\mathbb{E}\left [ {\textstyle \sum_{k=1}^{K}}\left \|\tilde{x}_0-\hat{x}_k\right \|_1 \right ].
\label{eq:l_sel}
\end{equation}
where $\hat{x}_k$ refers to the view reconstructed from $\bar{x}_k$, $K=N \times (N-1)$.

Consequently, FMAug not only extends the domain discrepancy and margin, but also poses a self-supervised pretext task to learn generalizable context-aware representations from view reconstruction.

\subsection{Coupled segmentation network} 
As the FMAug promises domain-augmented training data and generalizable context-aware representations, a segmentation model capable of out-of-domain inference is waiting to be learned.
To inject the context-aware representations into the segmentation model seamlessly, a coupled network is designed with attention mechanisms (shown in the purple block of Fig.~\ref{fig:workflow}), which utilize the most relevant parts of representation in a flexible manner.


Concretely, the network comprises an encoder $E$ and two decoders $D_{sel}$, $D_{seg}$, where skip connection bridges $E$ and $D_{sel}$ while $D_{sel}$ marries $D_{seg}$ using attention mechanisms.
For the above pretext task, $E$ and $D_{sel}$ compose a U-Net architecture to reconstruct $\tilde{x}_0$ from $\bar{x}_k$ with the objective function given in Eq.~\ref{eq:l_sel}. 
On the other hand, the segmentation task shares $E$ with the pretext task, and introduces representations from $D_{sel}$ to $D_{seg}$.
The features outcomes from the $l$-th layer of $D_{seg}$ are given by
\begin{equation}
f^l_{seg}=D^l_{seg}([f^{l-1}_{seg},f^{l-1}_{sel}]),\;l=1,2,...,L,
\label{eq:fr}
\end{equation}
where $f^{l}_{sel}$ refers to the features from the $l$-th layer of $D_{sel}$. 
Additionally, attention modules are implemented to properly couple the features from $D_{sel}$ and $D_{seg}$.
$D^l_{seg}$ imports and concatenates $f^{l-1}_{seg}$ and $f^{l-1}_{sel}$ as a tensor.
Subsequently,  the efficient channel and spatial attention modules proposed by~\cite{woo2018cbam} are executed to couple the representations learned from the pretext and segmentation task.
Then convolutional layers are used to generate the final outcome $f^l_{seg}$.
Accordingly, denote the segmentation result from $\bar{x}_k$ as $\hat{m}_k$, the objective function for segmentation task is given by
\begin{equation}
\mathcal{L}_{seg}=\mathbb{E}\left [ {\textstyle \sum_{k=1}^{K}}[-m\log{\hat{m}_k}-(1-m)\log{(1-\hat{m}_k)}]\right ].
\label{eq:l_seg}
\end{equation}
where $m$ denotes the ground-truth segmentation mask corresponding to the original source sample $x$.
Therefore, the overall objective function for the network is defined as
\begin{equation}
\mathcal{L}_{total}=\mathcal{L}_{sel}(E,D_{sel})+\alpha \mathcal{L}_{seg}(E,D_{sel},D_{seg}),
\label{eq:overall}
\end{equation}
where $\alpha$ is the hyper-parameter to balance $\mathcal{L}_{sel}$ and $\mathcal{L}_{seg}$.

\section{Experiments}
\noindent \textbf{Implementation:} Five image datasets of three modalities were collected to conduct segmentation experiments on fundus vessels and articular cartilage. 
For fundus vessels, training was based on 1) DRIVE\footnote[1]{http://www.isi.uu.nl/Research/Databases/DRIVE/} and 2) EyePACS\footnote[2]{https://www.kaggle.com/c/diabetic-retinopathy-detection}, where DRIVE is a vessel segmentation dataset on fundus photography used as the single source domain to learn a generalizable segmentation model, EyePACS is a tremendous fundus photography dataset employed as extra multiple source domains to implement DG-based algorithms.
3) LES-AV\footnote[3]{https://figshare.com/articles/dataset/LES-AV\_dataset/11857698/1} and 4) IOSTAR\footnote[4]{http://www.retinacheck.org/datasets} are vessel segmentation datasets respectively on fundus photography and Scanning Laser Ophthalmoscopy (SLO), which were used to verify the generalizability of models learned from DRIVE.
For articular cartilage, 5) ultrasound images of joints with cartilage masks were collected by Southern University of Science and Technology Hospital, under disparate settings to validate the algorithm's effectiveness in multiple medical scenarios, where the training, generalization, and test splits respectively contain 517, 7530, 1828 images.

The image data were resized to $512\times512$, the training batch size was 2, and Adam optimizer was used. The model was trained according to an early-stop mechanism, which means the optimal parameter on the validation set was selected in the total 200 epochs, where the learning rate is 0.001 in the first 80 epochs and decreases linearly to 0 in the last 120 epochs.
The encoder and two decoders are constructed based on the U-net  architecture with 8 layers. 
The comparisons were conducted with the same setting and were quantified by DICE and Matthews's correlation coefficient (Mcc).

\vspace{3pt} \noindent \textbf{Comparison and Ablation Study:} The effectiveness of the proposed algorithm is demonstrated in comparison with state-of-the-art methods and an ablation study. 
The Fourior-based DG methods FACT~\cite{QinweiXu2021AFF}, FedDG~\cite{liu2021feddg}, and the whitening-based DG method SAN-SAW~\cite{peng2022semantic}, as well as the SGD method GIN-IPA~\cite{ouyang2022causality} were compared, where CE-Net~\cite{gu2019net} and CS-Net~\cite{mou2019cs} were served as the base models cooperated with FACT~\cite{QinweiXu2021AFF}.
Then in the ablation study, FMAug,  self-supervised learning (SSL), and attention mechanisms (ATT) were respectively removed from the proposed algorithm.


\begin{figure}[!t]
    \centering
        \includegraphics[width=0.9\linewidth]{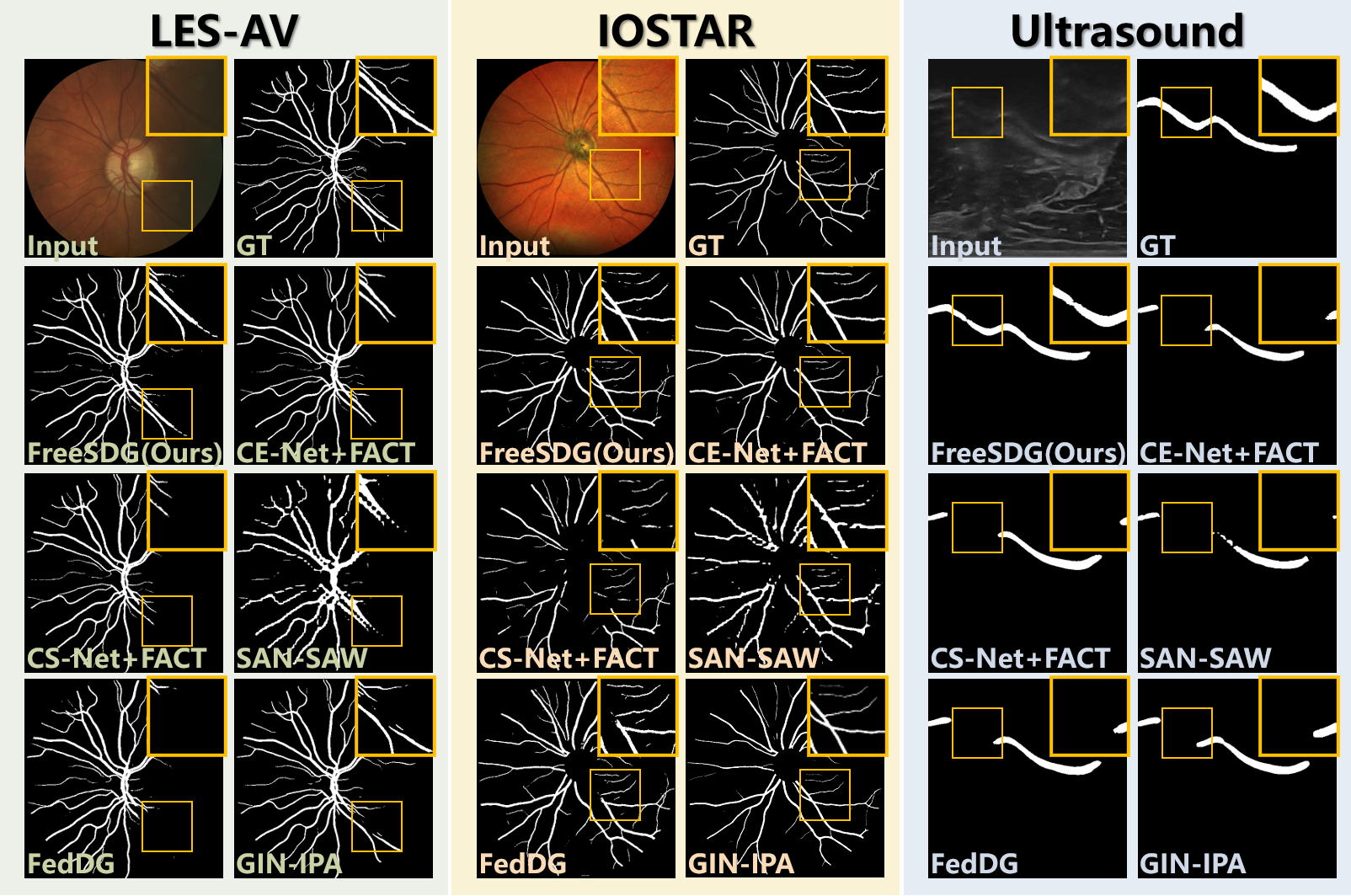} 
    \caption{Segmentation comparison in three medical image modalities.} 
    \vskip -5pt
    \label{fig:output}
\end{figure}

\vspace{3pt} \noindent \textbf{(1) Comparison}
Quantified comparison of our algorithm with the competing methods is summarized in Table~\ref{tab:comparison}, where segmentation results in three modalities and data dependency are exhibited.
Due to the domain shifts between DRIVE and LES-AV as well as IOSTAR, interior performance are presented by CE-Net~\cite{gu2019net} and CS-Net~\cite{mou2019cs}, which are only learned from DRIVE without DG.
Due to the substantial domain discrepancy, EyePACS were treated as multiple source domains to implement DG.
FACT~\cite{QinweiXu2021AFF} boosts the generalization by transferring LFS across the multi-source domains, and efficiently promotes the performance of CE-Net~\cite{gu2019net} and CS-Net~\cite{mou2019cs}.
FedDG~\cite{liu2021feddg} were then respectively trained using DRIVE perturbed by EyePACS.
As SAN-SAW~\cite{peng2022semantic} was designed for region structure segmentation, it appears redundant in the vessel structure task.
Thanks to coupling federated learning and contrastive learning, reasonable performance are provided by FedDG~\cite{liu2021feddg}.
GIN-IPA~\cite{ouyang2022causality} and our FreeSDG were learned based on the single source domain of DRIVE.
Through augmenting the source domain with intensity variance and consistency constraint, GIN-IPA~\cite{ouyang2022causality} performs decently on out-of-domain inference. 
The proposed FreeSDG allows for learning efficient segmentation models only from DRIVE.
Therefore, our FreeSDG outperforms the state-of-the-art methods without extra data dependency.
Additionally, an identical situation is observed from the results of ultrasound data, further validating the effectiveness of our algorithm.

Visualized comparison is shown in Fig.~\ref{fig:output}.
Uneven brightness in LES-AV impacts the segmentation performance, vessels in the highlight box are ignored by most algorithms.
Cooperating with FACT~\cite{QinweiXu2021AFF}, CE-Net~\cite{gu2019net} achieves impressive performance.
The remarkable performance of GIN-IPA~\cite{ouyang2022causality} indicates that SDG is a promising paradigm for generalizable segmentation.
In the cross-modality segmentation in IOSTAR, CE-Net~\cite{gu2019net} married with FACT~\cite{QinweiXu2021AFF}  and GIN-IPA~\cite{ouyang2022causality} still performs outstandingly.
In addition, decent segmentation is also observed from FedDG~\cite{liu2021feddg} via DG with multi-source domains.
FreeSDG efficiently recognizes the variational vessels in LES-AV and IOSTAR, indicating its robustness and generalizability in the quantitative comparison.
Furthermore, FreeSDG outperforms the competing methods in accurately segmenting low-contrast cartilage of ultrasound images.
In nutshell, our SDG strategy promises FreeSDG prominent performance without extra data dependency.

\begin{table}[!t]
\scriptsize
\centering
\caption{Comparisons and ablation study}
\label{tab:comparison} 
\renewcommand{\arraystretch}{1.15}
\begin{threeparttable}
\begin{tabular}{p{1.9cm} | | p{0.9cm}<{\centering} p{0.9cm}<{\centering} | | p{1.1cm}<{\centering} p{1.1cm}<{\centering} |p{1.1cm}<{\centering}  p{1.1cm}<{\centering} |p{1.1cm}<{\centering} p{1.1cm}<{\centering} }
\hline
\multirow{2}{*}{Algorithms} &\multicolumn{2}{c||}{\textsl{Dependency}*}& \multicolumn{2}{c|}{LES-AV} & \multicolumn{2}{c|}{IOSTAR} & \multicolumn{2}{c}{Ultrasound} \\
\cline{2-9}
& \textsl{IID} & \textsl{MSD} &  DICE   &  Mcc  &  DICE  &  Mcc  &  DICE  &  Mcc  \\
\hline
CE-Net &\textcolor{red}{$\star$}&& 0.636 & 0.618 & 0.505 & 0.514 & 0.788 & 0.796 \\
CS-Net &\textcolor{red}{$\star$}&& 0.593 & 0.559 & 0.520 & 0.521 & 0.699 & 0.721 \\
CE-Net+FACT &&\textcolor{red}{$\star$}& 0.730 & 0.711 & 0.728 & 0.705 & 0.846 & 0.846 \\
CS-Net+FACT &&\textcolor{red}{$\star$}& 0.725 & 0.705 & 0.580 & 0.572 & 0.829 & 0.827 \\
SAN-SAW &&\textcolor{red}{$\star$}& 0.629 & 0.599 & 0.617 & 0.585 & 0.819 & 0.822 \\
Feddg &&\textcolor{red}{$\star$}& 0.745 & 0.725 & 0.720 & 0.697 & 0.872 & 0.871 \\
GIN-IPA &&& 0.683 & 0.665 & 0.641 & 0.650 & 0.827 & 0.824 \\
FreeSDG(ours) &&& \textbf{0.795} & \textbf{0.778} & \textbf{0.736} & \textbf{0.716} & \textbf{0.913} & \textbf{0.912} \\
\hline
\multicolumn{3}{l||}{FreeSDG w/o FMAug, SSL, ATT} & 0.720 & 0.705 & 0.687 
 & 0.665 & 0.875 & 0.873 \\
\multicolumn{3}{l||}{FreeSDG w/o SSL, ATT} & 0.751 & 0.734 & 0.724 & 0.701 & 0.881 & 0.881 \\
\multicolumn{3}{l||}{FreeSDG w/o ATT} & 0.777 & 0.760 & 0.731 & 0.709 & 0.898 & 0.897 \\
\hline
\end{tabular}%
\begin{tablenotes}
 \scriptsize
 \item{*} Independent and identically distributed data (\textsl{IID}) and multi-source domains (\textsl{MSD}).
\end{tablenotes}
\end{threeparttable}
\vskip -5pt
\end{table}

\vspace{3pt} \noindent \textbf{(2) Ablation Study}
According to Table~\ref{tab:comparison}, the ablation study also validates the effectiveness of the three designed modules.
Through FMAug, an augmented source domain with adequate discrepancy is constructed for training generalizable models.
Robust context-aware representations are extracted from self-supervised learning, boosting the downstream segmentation task.
Attention mechanisms seamlessly inject the context-aware representations into segmentation, further improving the proposed algorithm.
Therefore, a promising segmentation model for medical images is learned from a single-source domain.

\section{Conclusion}
Pixel-accurate annotations have long been a common bottleneck for developing medical image segmentation networks.  
Segmentation models learned from a single-source dataset always suffer performance dropping on out-of-domain data. 
Leveraging DG solutions bring extra data dependency, limiting the deployment of segmentation models.
In this paper, we proposed a novel SDG strategy called FreeSDG that leverages a frequency-based domain augmentation technique to extend the single-source domain discrepancy and injects robust representations learned from self-supervision into the network to boost segmentation performance. Our experimental results demonstrated that the proposed algorithm outperforms state-of-the-art methods without requiring extra data dependencies, providing a promising solution for developing accurate and generalizable medical image segmentation models.
Overall, our approach enables the development of accurate and generalizable segmentation models from a single-source dataset, presenting the potential to be deployed in real-world clinical scenarios.

\section*{Acknowledgment}
This work was supported in part by Basic and Applied Fundamental Research Foundation of Guangdong Province (2020A1515110286), the National Natural Science Foundation of China (82102189, 82272086), Guangdong Provincial Department of Education (2020ZDZX3043), Guangdong Provincial Key Laboratory (2020B121201001), Shenzhen Natural Science Fund (JCYJ20200109140820699, 20200925174052004), Shenzhen Science and Technology Program (SGDX202111 23114204007), Agency for Science, Technology and Research (A*STAR) Advanced Manufacturing and Engineering (AME) Programmatic Fund (A20H4b0141) and Central Research Fund (CRF).
%
%

\bibliographystyle{splncs04}
\bibliography{briefbib}


%




\end{document}